\documentclass[traditabstract]{aa}  
\usepackage[colorlinks=true,linkcolor=bluea,allcolors=blue]{hyperref}
\usepackage{graphicx}
\usepackage{txfonts}

\makeatletter
\renewcommand*\aa@pageof{, page \thepage{} of \pageref*{LastPage}}
\makeatother

\newcommand{\lya}{Ly$\alpha$}

\newcommand{\kms}{km s$^{-1}$}

\newcommand{\TXS}{TXS~0952-217}
\newcommand{\TN}{TN~J1049-1258}
\usepackage[switch]{lineno}
\begin{document}

   \title{A 20 kiloparsec bipolar Lyman $\alpha$ outflow from a radio galaxy at z=2.95.}

   \author{M. Coloma Puga\inst{1,2}
          \and
          B. Balmaverde\inst{2}
          \and
          A. Capetti\inst{2}
          \and
          C. Ramos Almeida\inst{3,4}
          \and
          F. Massaro\inst{1}
          \and
          G. Venturi\inst{5}
        }
   \institute{Department of Physics \\
            Università di Torino \\
            Via Pietro Giuria, 1, 10125, Torino, Italy
            \and
            INAF - Osservatorio Astrofisico di Torino \\
            Via Osservatorio 20, I-10025 Pino Torinese, Italy
            \and
            Instituto de Astrofisica de Canarias\\
            C. Vía Láctea, s/n, 38205 La Laguna, Santa Cruz de Tenerife, Spain    
            \and
            Departamento de Astrofísica, Universidad de La Laguna \\
            38206, La Laguna, Tenerife, Spain
            \and
            Scuola Normale Superiore \\
            P.za dei Cavalieri, 7, 56126 Pisa PI
            }

   \date{Received September 15, 1996; accepted March 16, 1997}

% \abstract{}{}{}{}{} 
% 5 {} token are mandatory
 
  \abstract{ 
   The study of ionized gas kinematics in high-z active galaxies plays a key part in our understanding of galactic evolution, in an age where nuclear activity was widespread and star formation close to its peak. We present a study of \TXS, a radio galaxy at z=2.95, using VLT/MUSE integral field optical spectroscopy as part of a project aimed studying of the properties of ionized gas in high redshift radio galaxies (HzRGs). The \lya\ line profile of this object presents various emission and absorption components. By utilizing Voronoi binning, we obtained a comprehensive map of the kinematic properties of these components. These observations revealed the presence of a redshifted, high velocity (v $\sim 500$ \kms) bipolar structure of \lya\ emission, most likely corresponding to an outflow of ionized gas. The outflow extends beyond the compact radio source on both sides, with a total size of $\sim$ 21 kpc. Its kinetic power ($10^{42.1}$ erg s$^{-1}$) is about five orders of magnitude smaller than its radio power. Additional ionized lines, including HeII$\lambda$1640, CIV$\lambda$1550 and CIII]$\lambda$1908 were detected and their line flux ratios determined. The presence of HeII allowed for a precise redshift measurement (z=2.945$\pm$0.002). Along with the recent discovery of a similar structure in \TN, another HzRG, it displays the feasibility of using \lya\ as a tracer of outflowing gas in high redshift sources, and particularly so when supported by non-resonant ionized lines such as HeII, which allow for accurate redshift and velocity measurements.}


   \maketitle
%
%-------------------------------------------------------------------

\section{Introduction}

The majority of \lya\ (1215.67\r{A}) radiation in galaxies is emitted by ionized hydrogen atoms undergoing recombination, as the electron cascades down to lower energy levels \citep{Cantalupo_2017}. Among the possible sources of ionization are both active galactic nuclei (AGN) and populations of young, massive stars. Consequently, star-forming galaxies that host nuclear activity are some of the brightest \lya\ emitters. The \lya\ emission line is now routinely utilized to study high-redshift objects as it falls in the optical band, allowing for deep ground-based observations, either through spectroscopy or narrowband imaging.

\lya\ has also been observed to form very large extended structures around active high redshift galaxies \citep{Cantalupo_2014,borisova16}, with sizes of up to hundreds of kiloparsecs, in both diffuse halos and cosmic-scale filaments of gas. Photons emitted in \lya\, originated from the dense ionized gas (\citealt{Prochaska_2013}, \citealt{Finley_2014}) surrounding these active galaxies, enable the observation of these gargantuan structures through fluorescent emission, probing deeper than other lines thanks to its intrinsic brightness \citep{Hayes_2015}.

Integral field units, and in particular the Multi Unit Spectroscopic Explorer (MUSE) at the Very Large Telescope (VLT) and the Keck Cosmic Web Imager (KCWI) at Keck II, have enabled the systematic search and characterization of these high-redshift \lya\ nebulae around AGN, from z$\sim$2 onward in the case of KCWI (e.g., \citealt{Cai_2019}) and from z$\sim$3 in the case of MUSE (e.g., \citealt{Arrigoni_Battaia_2018}). The nebulae appear to be brighter and more extended at higher redshifts, possibly due to a decrease in the cool-to-warm ionized gas mass ratio with cosmic age.
 
In galaxies themselves, however, the physical processes underlying this emission result in a complex line profile that is difficult to fully interpret in terms of its relation to the physical and kinematic state of the emitting gas. \lya\ photons are scattered due to both the large cross-section of interaction with neutral hydrogen and the ubiquity of this element, increasing the optical path length these photons must travel to escape the galaxies they originate from. As the path length increases, so do the chances of these photons interacting with dust grains, which will re-emit this energy in the infrared range and strongly influence the amount of \lya\ radiation that can escape from a given galaxy.

As a result of the aforementioned resonance, the outgoing \lya\ emission has a complex line profile usually characterized by a bright emission component and fainter bumps at shorter wavelengths. This asymmetry affects the centroid of the line, resulting in a displacement of the measured central wavelength and discrepancies with other nonresonant ionized lines: redshift measurements based on \lya\ have been found to be offset by an average of $\sim$400 \kms\ when compared to  nebular lines \citep{steidel2010}. Additionally, gas kinematics such as those stemming from outflowing winds can yield line profiles with multiple peaks and absorption troughs (see \citealt{Verhamme_2006}, \citealt{Behrens_2014}, \citealt{Gronke_2015} and \citealt{Gronke_2017} and references therein for various models and numerical simulations of the radiative transfer of \lya).

Nonetheless, the \lya\ emission properties in high-redshift AGN have been a topic of recent study, with deep integral field observations enabling the detection of large-scale ionized gas kinematic signatures originating from outflowing gas, accretion flows, and cool HI absorbers, among other phenomena \citep{Swinbank_2015,Silva_2017,Arrigoni_Battaia_2019,Travascio_2020,Wang_2021,Lau_2022,Wang_2023,wang2024jwst}. These studies have observed the effect AGN have on the \lya\ nebulae surrounding their host galaxies, as well as the enrichment of the intergalactic medium traced by metallic emission and absorption lines, such as HeII$\lambda$1640 and CIV$\lambda$1550.

One of the strongest pieces of evidence highlighting the influence AGN can have on their host galaxies are galactic-scale outflows, driven by radiation pressure and/or relativistic jets. They have been postulated as the mechanism through which active galaxies regulate star formation and interact with their environment (feedback). It has been hypothesized that all galaxies undergo short, recurring phases of nuclear activity (e.g., \citealt{Schawinski_2015}) that lead to the coevolution of supermassive black holes and their host galaxies \citep{Kormendy_2013}. Moreover, cosmological simulations lacking AGN feedback produce an excess of massive, star-forming galaxies (\citealt{Dubois_2016}). Consequently, feedback from AGN due to jet- or radiation-induced outflows (kinetic or radiative feedback) is one of the most important factors in explaining and predicting the observed properties of massive galaxies. 

While the reasons and evidence outlined above indicate both the importance of feedback in simulations and the necessity of its existence as inferred from the observed properties of galaxies, direct observations have not resulted in a clear picture of how AGN feedback operates across cosmic time and how effective it actually is in affecting the evolution and star formation activity in galaxies. There are a few isolated cases with evidence of feedback in action in the form of outflows quenching star formation (e.g. \citealt{Cresci_2015}, \citealt{Carniani_2016}), but the effect seems to be limited to some parts of the galaxy and unable to affect the star formation rate as a whole. Moreover, these pieces of evidence are still debated (see \citealt{Scholtz_2020,Scholtz_2021}).

On the other hand, some of the clearest signs of AGN feedback in action come from radio galaxies (RGs). In the massive, brightest cluster galaxies in the local Universe, jets are observed to mechanically inject energy into the hot, X-ray-emitting halo, preventing gas accretion and thus star formation (see e.g. \citealt{Best_2006,Balmaverde_2018}).

High-redshift radio galaxies (HzRGs) are the progenitors of such massive galaxies, and we are able to observe them close to or at cosmic noon (z$\sim$2-3), an age of the Universe when galaxies were still evolving and forming stars at high rates \citep{Madau2014}. The consequences of AGN feedback, postulated as a mechanism through which galaxies regulate their own growth and star formation, should be particularly pronounced in these sources. Along with type 2 quasars, RGs are characterized by relatively low continuum nuclear luminosities, since the dusty torus (and/or the galactic component itself) situated along our line of sight prevents the accretion disk from outshining the host galaxy \citep{almeida2017nuclear}. Consequently, the bright point spread function that has to be accounted for when observing AGN face-on is not present, making it possible for the galactic emission to be spatially resolved and studied at the smallest scales permitted by instrumentation.

Given both the intrinsic brightness and resonant properties of \lya, rest-frame UV observations of HzRGs allow us to study the distribution and kinematics of ionized and neutral hydrogen in their environment. These observations enable a full characterization of these sources, as one can observe massive, galactic-scale outflows in emission and inflows from the cosmic web and nearby gas in absorption.

HzRGs have been the subject of many studies (see \citealt{Miley_2008} for a review, and references therein), as they tend to be particularly massive \citep{De_Breuck_2010} and consequently, bright. RGs have been observed to follow a tight K-band magnitude-to-redshift relation \citep{willott99,brookes06,Seymour_2007}, beginning in the local Universe and up to z$\sim$4, with their estimated stellar masses remaining fairly constant across cosmic time, up to an observed "hard" limit of $10^{12} M_{\odot}$ \citep{Rocca_Volmerange_2004}. Of particular interest to our work are the early studies of \lya\ emission around HzRGs \citep{vanojik1996radio,vanojik1996gaseous,Reuland_2003,Villar_Mart_n_2007_a,Villar_Martin_2007_b}. Recent work, such as the study performed by \citealt{Wang_2023} highlights the possibilities large-scale surveys of this \lya\ emission have to offer.

Also of interest are the environments inhabited by these galaxies; low-redshift RGs have also been found to reside in particularly dense environments and galaxy clusters \citep{Donoso_2010,Falder_2010}, in many cases as the brightest cluster galaxy \citep{Hogan_2015,Kale_2015}. HzRGs have been postulated as the progenitors of these massive radio-loud galaxies, as they too generally reside within overdense environments \citep{Venemans_2006,Wylezalek_2013,Rigby_2013,Uchiyama_2022} that, while not yet tied by gravity at this early cosmic age, will in many cases become bound and form clusters.

However, we still do not know how the evolution of HzRGs occurs and how much and with which mechanisms an AGN affects the host galaxy in this class of object. It is thus necessary to gain a detailed understanding of the effects of AGN-related phenomena on the interstellar medium and on the star formation history, in the framework of galactic evolution.

We began a comprehensive program of observations with MUSE at the VLT to explore the properties of the ionized gas in HzRGs. The capabilities of this instrument combine the simultaneous spatial and spectral information of an integral field unit with the high sensitivity of an 8 m class telescope, allowing for detailed studies of the properties of \lya\ emitters at z$\geq$2.9. In \citet{puga2023extended} we presented the results obtained for \TN, a HzRG at z=3.7, for which the MUSE data show the presence of a one-sided, high-velocity ($\sim 2,250$ \kms) \lya\ structure extending for $\sim 18$ kpc along the radio jet, most likely an outflow of ionized gas. 

In this work we present the results obtained for the central regions of the \lya\ nebula in TXS~0952-217, an HzRG at a redshift of 2.95 \citep{Brookes_2008} located at RA=09:54:28.9, Dec=-21:56:53.6. The scale factor at this redshift (assuming an H$_{0}$=69.6\kms Mpc$^{-1}$ and $\Omega_m$=0.286 cosmology) is 7.9 kpc arcsec$^{-1}$. 

\section{Observations and data analysis}\label{sec:obs}

The observations were carried out on the 10 January 2022 as part of the program ID:108.22FU (PI:Balmaverde). Four separate observations were performed, between which the telescope was rotated 90$^\circ$ to reject cosmic rays, for a total of 2800s of exposure time. We used the European Southern Observatory (ESO) MUSE pipeline (version 2.8.7) to obtain a fully reduced and calibrated data cube \citep{Weilbacher20}. The seeing was measured to be 0\farcs72$\pm$0\farcs03 from various stars within the field. The only galaxy within the 1x1 arcmin$^{2}$ field of view that displays \lya\ emission is the host of \TXS.

\subsection{Source information}\label{subsec:source}

\TXS\ is unresolved in all available radio observations. The stronger constraint on its size is provided by the \textit{Karl G. Jansky} VLA Sky Survey (VLASS, \citealt{Lacy_2020}), performed at 3 GHz with a resolution of 2\farcs5, where it appears as a compact unresolved source with an upper limit to its deconvolved size of $\sim1\farcs2$ corresponding to $\sim$10 kpc. Its radio spectrum, obtained by combining measurements from the Galactic and Extra-Galactic All-Sky MWA (GLEAM) catalog \citep{hurley17} with those from VLASS, is well described by a single power law defined as a flux density $F_\nu \propto \nu^{-1}$. The jet power, $P_{\rm jet}$, of \TXS\ can be estimated using the empirical relation between $P_{\rm jet}$ and radio luminosity at 327 MHz, $P_{327}$ \citep{Cavagnolo10}. The radio luminosity of \TXS\ at (rest frame) 327 MHz, extrapolated from the observed GLEAM flux density at 92 MHz, is $P_{327} = 3.4\times10^{44}$ erg s$^{-1}$ which yields a jet kinetic power of $P_{\rm jet} \sim 10^{47}$ erg s$^{-1}$. From \citealt{Miley_2008}, the radio luminosity of \TXS\ is at the lower end of those observed for HzRGs.

The K-band magnitude of \TXS\ is K=19.87$\pm$0.20 mag as measured from the United Kingdom Infrared Telescope (UKIRT) fast-track imager\citep{brookes06}. The observed K-band luminosity, which approximately corresponds to the V band in the rest frame, is $\sim 10^{11} {\rm L}_\odot$. This value is in line with the expected stellar mass of HzRGs \citep{Rocca_Volmerange_2004,Seymour_2007}.

The lack of suitable sources prevented us from deriving an accurate astrometric correction for the MUSE data and consequently, we are unable to give an approximate position for the radio source with respect to these data. We assumed that the unresolved radio source is coincident with the peak of the continuum emission, which itself is coincident with the peak of \lya.

\subsection{Redshift determination}\label{subsec:red}

\begin{figure}[ht]
    \centering
    \includegraphics[width=0.5\textwidth]{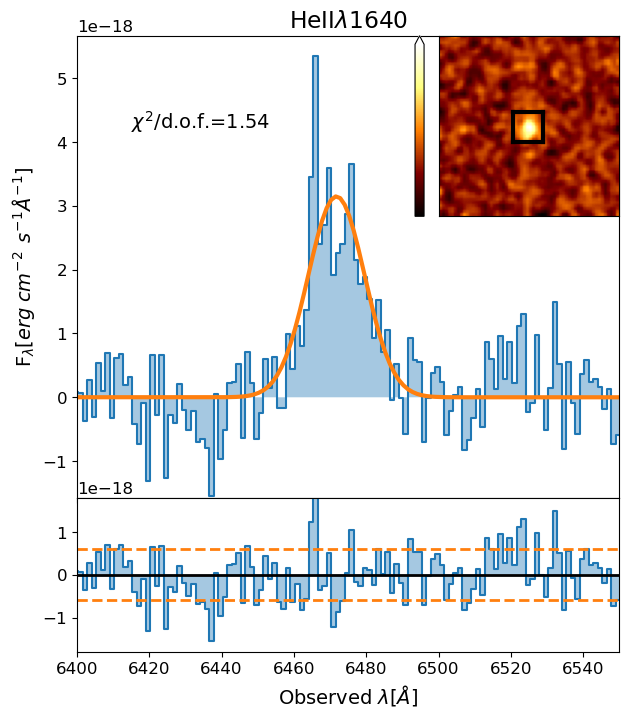}
    \caption{\textbf{Top.} Continuum-subtracted HeII$\lambda$1640 emission line profile extracted from a $2\farcs0\times2\farcs0$ rectangular aperture centered on the emission peak. The position of the aperture relative to the HeII emission, extracted using a 20\r{A} wide band is displayed in the top right inset. The orange line corresponds to the modeled emission profile. \textbf{Bottom.} Residuals from the fit, with the dashed orange lines marking the measured noise level.}
    \label{fig:heiifit}
\end{figure}

Prior to the analysis of the \lya\ properties, the HeII$\lambda$1640 emission line was fit (see Figure \ref{fig:heiifit}), to obtain the redshift of the host galaxy and use it to constrain the center of the intrinsic \lya\ emission. This was done because the measured centroid of nonresonant lines is a reliable tracer of systemic redshift, while the observed \lya\ center has a tendency to be offset by an average of $\sim$500 \kms redward of other nebular lines \citep{steidel2010}, another consequence of the asymmetrical absorption that results from resonance.

The redshift measured from the HeII$\lambda$1640 emission is z=2.945$\pm$0.002 ($\lambda_{\rm HeII}$=6471.3$\pm$1.6\r{A}). This is slightly lower than the value of z=2.95 reported by \citet{Brookes_2008} from \lya\, corresponding to a velocity shift of $\sim$360 \kms redward of the systemic redshift, which is within expectations \citep{steidel2010}. The fit residuals shown in Figure \ref{fig:heiifit} hint at a double peaked profile with two very narrow components, and this solution does yield a lower value of the reduced $\chi^{2}$ statistical, but the resulting modeled emission clashes with the observed \lya\ profile, both in the width and position of the components. As such, we conclude that the observed narrow peaks were spurious, a result of random noise. HeII, along with the other detected lines barring \lya\, appear more compact than the extended emission nebula.

\subsection{Lyman $\alpha$ fitting}\label{subsec:lya}

\begin{figure}[ht]
    \centering
    \includegraphics[width=0.5\textwidth]{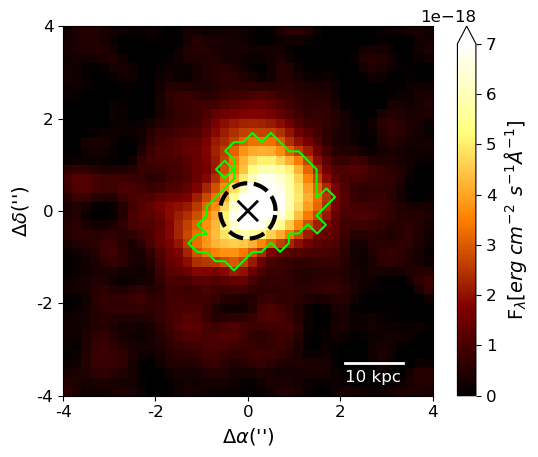}
    \caption{\lya\ narrowband image extracted using a 25\r{A} wide band centered on the approximate centroid of the \lya\ emission. The cross represents the brightest pixel, which is the assumed AGN position, and corresponds to the $\Delta\alpha$=0'',$\Delta\delta$=0'' position. The dashed circle represents the upper limit on the size of the radio emission under the same assumption of AGN position. The green contours represent the 2$\sigma$ detection limits. The image has been smoothed with a Gaussian 2D kernel.}
    \label{fig:lyanarrow}
\end{figure}

With the systemic redshift now determined, we began the \lya\ analysis process. We initially obtained a \lya\ image of the source producing a narrowband image extracted using a 25\r{A} wide band centered on the centroid of the \lya\ emission, shown in Fig. \ref{fig:lyanarrow}. It shows a bright central region with an elongation of $\sim 24$ kpc at a 2$\sigma$ limit in the SE-NW direction. This elongation, as well as any size quoted from hereon, is obtained by subtracting the seeing size in quadrature to the measured extension.

\begin{figure}[ht]
    \centering
    \includegraphics[width=0.5\textwidth]{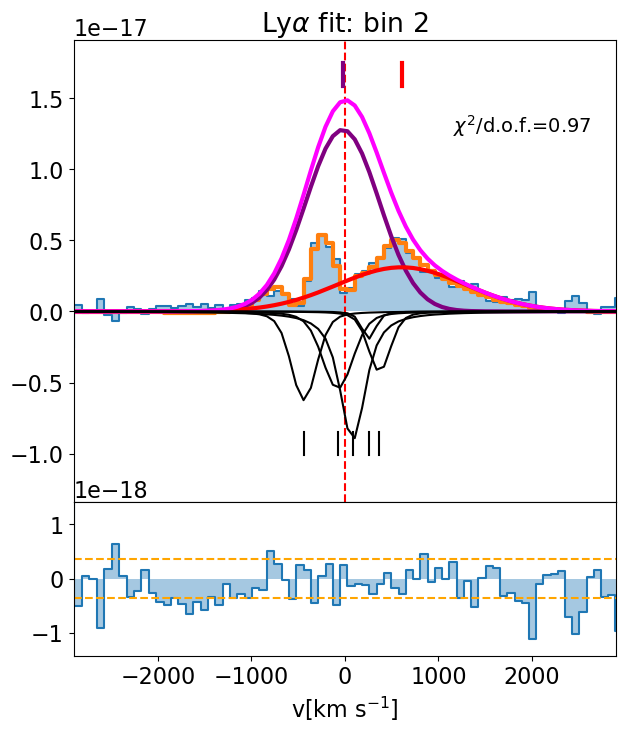}
    \caption{\textbf{Top:} Example spectrum extracted from one of the selected Voronoi bins. The blue shaded histogram represents the observed flux. Shown in purple is the main \lya\ component, in red the redshifted emission, in magenta the total combined emission and in black the absorption components. The vertical red and purple markings denote the emission component centroids. The orange histogram represents the total model, combining emission and absorption. \textbf{Bottom:} Fit residuals presented in the blue shaded histogram. The horizontal dashed orange lines represent the noise thresholds.}
    \label{fig:specexample}
\end{figure}

The \lya\ spectrum extracted in one of the optimally selected Voronoi regions within the nebula is shown in Fig. \ref{fig:specexample}; it corresponds to the sum of 14 spaxels (corresponding to 0.56 arcsec$^{2}$) located in the southeastern part of the nebula. As mentioned in the introduction, resonance can result in \lya\ emission yielding complex and varying profiles and indeed it shows multiple peaks and absorption features, the most prominent of which is at the source systemic redshift. The simultaneous fitting of emission and absorption profiles to \lya\ emission is a well-known procedure (see \citealt{vanojik1996gaseous,Jarvis_2003}), one that we adopt given the apparent complexity of the observed spectra.  We detect the presence of a high-velocity tail on the red side of the emission, visible up to $\sim$1800 \kms from the line center. If the whole \lya\ emission were to be attributed to a single Gaussian component, it would require a much larger amplitude than observed and a full width at half maximum (FWHM) of at least $\sim$2800 \kms; reproducing the profile would also require multiple clouds of absorbing gas moving at very high-redshifted velocities. Alternatively, adding an emission component with a velocity offset of about $\sim$+500 \kms\ from the line center reproduces the observed red tail and although absorbing systems close to the systemic velocity are still required, it eliminates the need for the extremely fast absorbers. Consequently, we consider the double-Gaussian plus the "lower" velocity absorbers solution to be the optimal method of analysis. There is precedent for modeling the intrinsic \lya\ emission with two Gaussians (see \citealt{Wang_2023}). The implications derived from the presence of such an emission component are discussed in later sections.

The \lya\ line profile is then fit with two emission components and several absorption troughs in different regions of the nebula. To maintain a relatively constant signal-to-noise ratio irrespective of position within the nebula, we divided the visible emission using Voronoi binning \citep{Cappellari2003}. The optimal compromise between S/N and spatial resolution is very much case-dependent, and given the apparent complexity of this source's profile, we settled on an S/N$\sim$8 over the full profile, which corresponds to a total of six bins. The decision on the S/N threshold was based on the resulting fits at different S/N values: lower levels of signal compromise the quality of the fitting, while the spatial resolution is not significantly improved. To avoid adding in pixels with low signal, we masked all pixels with individual S/N$<$1.5. We settle on this threshold based on a visual inspection of the remaining pixels after masking, as a higher cutoff eliminates slightly lower S/N pixels that contain otherwise relevant spectral information, while a lower cutoff includes pixels detached from the nebula as a result of random noise, and adding them to the binned spectra would contaminate them. The fitting requires between five and six absorbing profiles and two emission profiles. This degree of complexity demands a considerable amount of signal to successfully and reliably reproduce the data through modeling. We forced the center of the main \lya\ emission to be within $\pm$1\r{A} of the theoretical central wavelength corresponding to the redshift measured for the HeII line.

After identifying the emission components, absorption troughs modeled using Voigt profiles are added until our total model matches the observed profile. We used the value of the reduced $\chi^{2}$ as measure of goodness of fit. It is worth noting the used $\chi^{2}$ test exclusively evaluates how closely our model matches the observations, but it does not necesarily confirm the assumptions made to arrive at such a model (see \citealt{andrae2010dos}). The analysis is performed using the Python package \textit{specutils}\citep{astropy:2013,astropy:2018}. The Gaussian and Voigt models are fit with a Levenberg-Marquardt least squares method while the continuum is fit with a linear least squares method. The uncertainties that result from the fitting process are estimated through a Monte Carlo simulation. The fit is iterated (100 times for each bin in our case) while adding random noise to the spectrum; the noise is extracted from a normal distribution centered around the measured spectral noise for the selected bin. This results in a distribution of parameter values, from which we extracted both the parameter value itself and the 1$\sigma$ uncertainty.

\section{Results} \label{sec:res}

\begin{figure*}[ht]
    \centering
    \includegraphics[width=\textwidth]{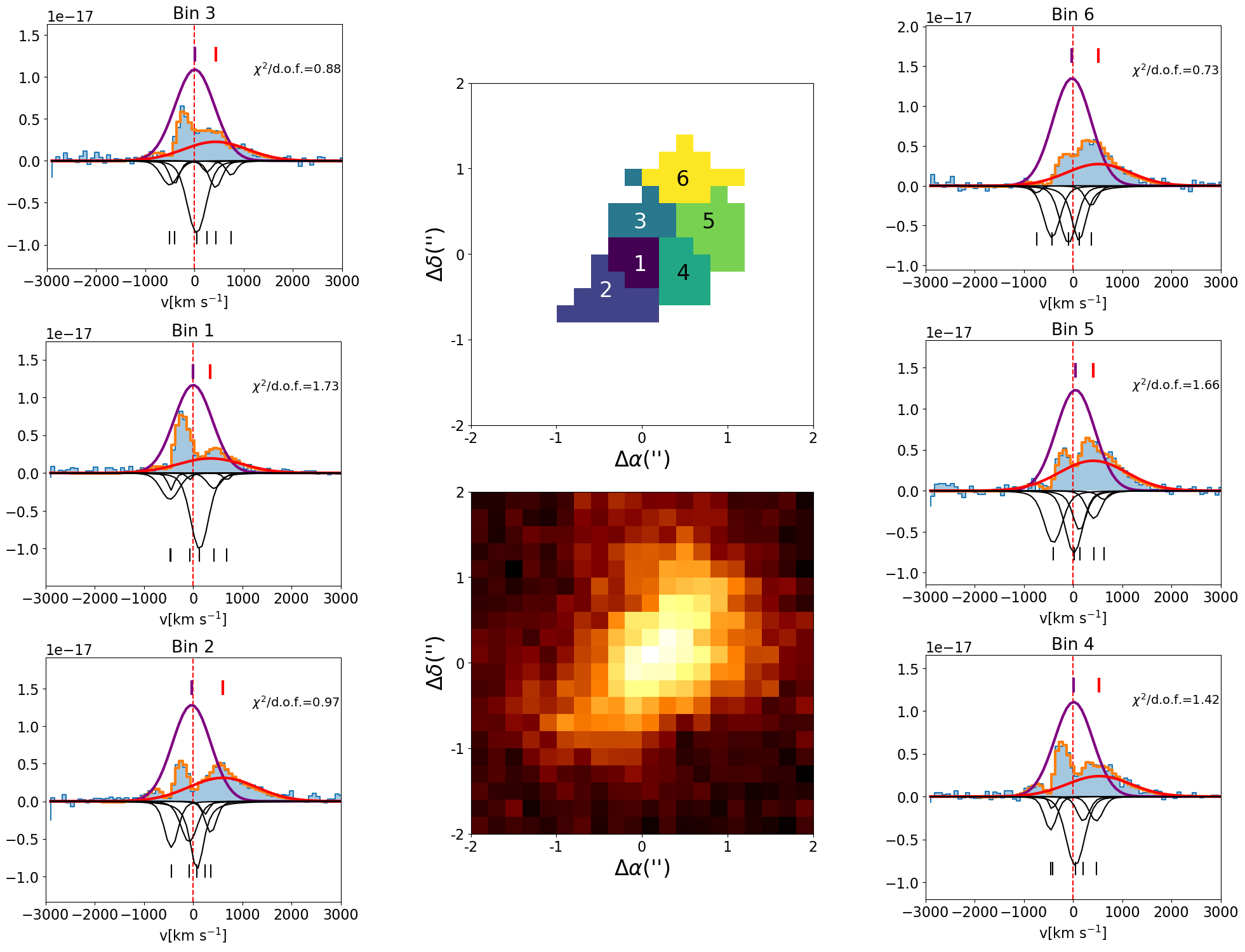}
    \caption{Profile analysis of the \lya\ nebula of \TXS. \textbf{Center top:} Pixels selected for the analysis based on an S/N>1.5 threshold, divided into the optimized Voronoi bins. Inside each bin is included an identifying number, which is referenced in the spectrum title. \textbf{Center bottom:} Narrowband image from which the pixels were selected. \textbf{Left and right:} Individual spectra in the Voronoi bins, identified by the bin number atop each plot. The blue shaded histogram represents the observed flux, while the orange line represents the total model, combining emission and absorption. Each individual component is also overlaid, in purple and red for the emission and in black for the absorption. The purple and red tick marks at the top correspond to the central velocities of each emission component, while the black tick marks at the bottom are their equivalents in absorption. All fluxes are given in erg cm$^{-2}$ s$^{-1}$\r{A}$^{-1}$.}
    \label{allbins}
\end{figure*}

The fit for the different bins, displayed in Figure \ref{allbins}, results in the detection of two emission components across the entire \lya\ emission region, and either five or six absorption troughs depending on the bin position.

\subsection{Gas kinematics}

We find the fit to be optimal using between five and six absorbers, some of which are below a 3$\sigma$ detection in amplitude when compared to the measured noise. Consequently, our discussion is centered on the major absorbers, those that result in a considerable impact to the final spectrum. Three of these absorbing clouds are detected in every bin; one very close and slightly redward of the systemic velocity, one further to the red, at a velocity of $\sim500$ \kms\ and one in the blue side of the spectrum, at $\sim-500$ \kms. In bins 5 and 6 a very deep absorption appears slightly to the blue of the systemic velocity, which is reflected on the observed spectra (see Figure \ref{allbins}). This would imply the presence of an optically thicker cloud of cold ionized gas toward the northwestern region of the nebula. A similar absorber is observed in bin 2, but given its diametrically opposed location within the nebula, its origin is likely a different gas cloud.

Regarding the gas kinematics for the main emission component, it is centered at $\lambda_{\rm main}=4796.2\pm0.4\;$\r{A} (corresponding to a velocity of v$_{\rm main}=25\pm25$ \kms) and we measure a width of FWHM$_{\rm main}=910\pm120$ \kms, averaged over all bins. For the red emission component, the center is at $\lambda_{\rm red}=4803.7\pm1.7\;$\r{A}(v$_{\rm red}=500\pm100$ \kms) the measured width is FWHM$_{\rm red}=1620\pm180$ \kms, also averaged. The kinematics of the main emission remain almost constant throughout the extension of the source, both in terms of centroid and velocity dispersion. For the red emission component, the kinematic maps are shown in Figure 5. While both emission components appear considerably broad, the measured width at the 80$^{\rm th}$ flux percentile (v$_{10}$-v$_{90}$) is W$_{80}$=1320$\pm$60\kms\, which is consistent with observations of \lya\ in other HzRGs \citep{Wang_2023}.

\begin{figure}[ht]
    \centering
    \includegraphics[width=0.5\textwidth]{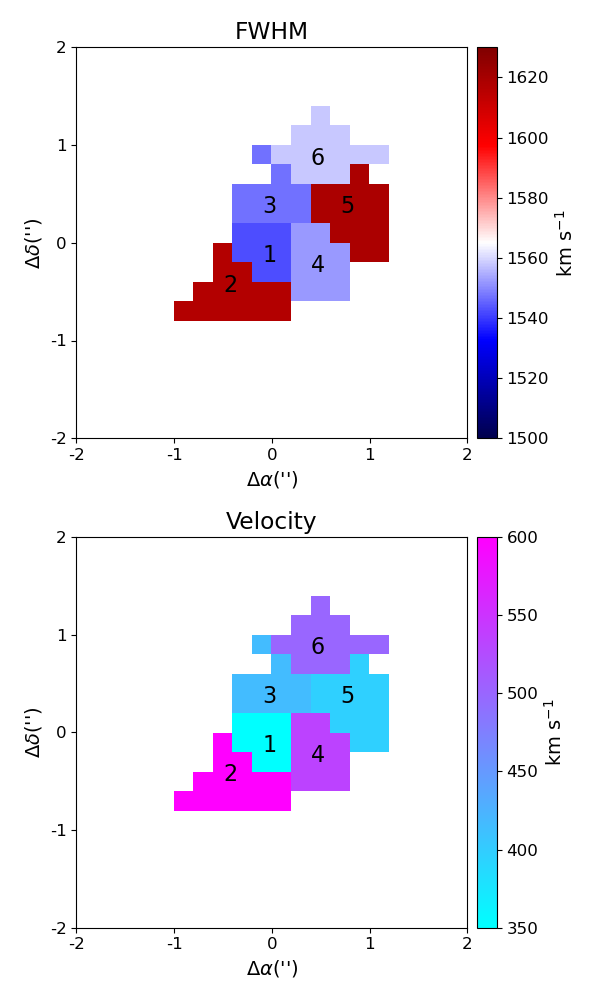}
    \caption{FWHM and velocity maps for the red emission component in each Voronoi bin.}
    \label{fig:velmaps}
\end{figure}

\begin{figure}[ht]
    \centering
    \includegraphics[width=0.5\textwidth]{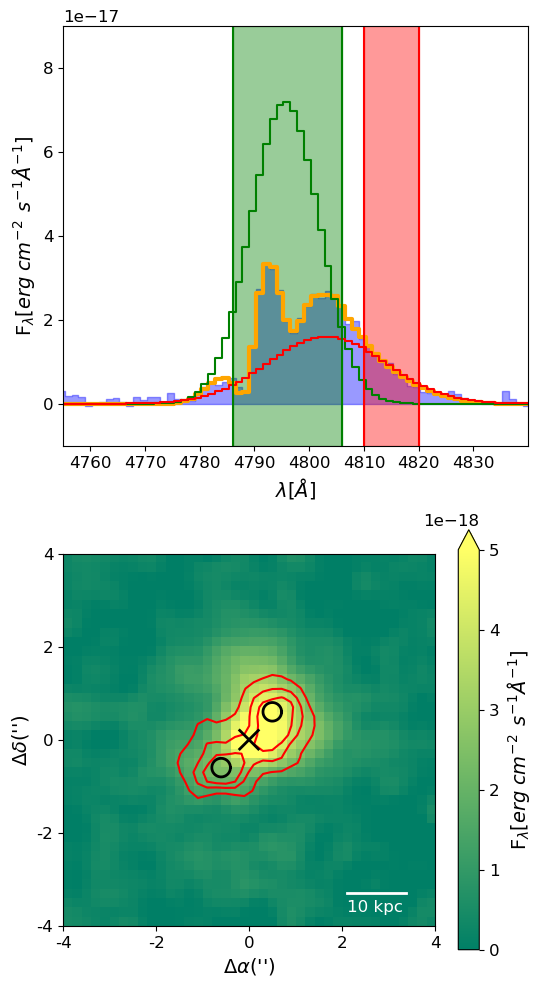}
    \caption{\textbf{Top:} Total spectrum extracted by summing all the Voronoi bins (shaded blue histogram) and the toal model, extracted in the same manner (shown in the orange histogram). The total emission modeled for the main and red components are shown in the green and red histograms. The green and red shaded areas correspond to the selected bands used for the extraction of the flux maps. \textbf{Bottom}: Color map of the \lya\ narrowband image extracted using a 20\r{A} wide band positioned on the approximate centroid of the \lya\ emission. The black cross marks the brightest pixel in \lya, which we assume coincides with the AGN position. The black circles mark the bright spots on each side of the red component. Shown in contours is the flux of the red tail from the \lya\ profile, extracted using a 10\r{A} band. The contour levels are set at 0.5, 0.8 and 1 $\times$ 10$^{-18}$ erg s$^{-1}$cm$^{-2}$.}
    \label{fig:lyatail}
\end{figure}

The identification of the presence of the main and of the high-velocity red wing enables us to perform narrowband imaging of the different spectral windows where each emission component is dominant. This, in turn, recovers some of the spatial information lost in the binning process and allows us to study the morphology of each component more closely. While both emission components are detected in all bins, the morphology derived from applying a synthetic narrowband to the red end of the emission profile is distinct from that of the main emission (Figure \ref{fig:lyatail}). The band position are width are chosen so as to avoid any absorption troughs while ensuring the contribution from the main component is negligible. When observed with this method, the red component shows a bipolar morphology, extending $\sim$2\farcs7 ($\sim$21 kpc) along the NW-SE direction, and on both sides of the AGN position, while the main component is more compact ($\sim$1\farcs8 or $\sim$14 kpc).

\subsection{Emission line ratios} 

We detect several other emission lines in the gas nebula, namely C~IV$\lambda$1550, He~II$\lambda$1640, C~III]$\lambda$1909 and N~V$\lambda$1240 (C~IV, He~II, C~III] and N~V from here on), the brightest lines being C~IV and He~II. With the exception of N~V, which is only detected in a few pixels, the other lines can be spatially resolved through narrowband imaging, using 20\r{A} wide bands centered on the HeII-measured redshift. Their emission is co-spatial with the main \lya\ emission (see Figure \ref{fig:lyatail}), but they do appear to be more compact, with sizes ranging from 1\farcs1 to 1\farcs4.

The line ratios are extracted by using continuum-subtracted narrowband fluxes in each of the selected bins. There are two reasons for following this procedure instead of fitting the lines and calculating the flux from the models. For the flux of the resonant lines (namely \lya\ and CIV$\lambda$1550), it enables comparisons with previous studies that also utilized narrowband measurements instead of fitting intrinsic emission and absorptions (e.g. \citealt{Villar_Mart_n_2007_a}). In the case of \lya\ in particular, the known degeneracy of the HI column density and amplitude of the intrinsic emission can result in uncertainties comparable to the magnitude of the measurement, which is wholly undesirable when estimating line ratios. For the other lines, it sidesteps the uncertainties that result from fitting the emission profile in bins where the signal-to-noise ratio is close to or below the threshold for a reliable detection.

In some cases, it is necessary to add the fluxes from adjacent bins in order to obtain a robust measurement of the line ratios. In particular, we combined the measurements for bins 5 and 6 into a single value, as well as bins 3 and 4. While this results in an additional loss of spatial resolution, we prioritized attaining a precise and reliable value. The line ratios obtained are presented in Table \ref{tab:en}.

\begin{table}[h]
\caption{Emission line ratios for the observed lines in \TXS. The regions utilized for the different measurements are given in the first column.}
    \centering
    \begin{tabular}{l | l l l l}
\hline        Bin N° & $\frac{\rm Ly\alpha}{{\rm He~II}}$ & $\frac{\rm Ly\alpha}{{\rm C~IV}}$ & $\frac{\rm C~IV}{\rm He~II}$ & $\frac{\rm C~IV}{\rm C~III]}$ \\ \hline
        1 & 5.6 $\pm$ 0.9 & 4.2 $\pm$ 0.8 & 1.3 $\pm$ 0.3 & 1.9 $\pm$ 0.6 \\
        2 & 5.3 $\pm$ 2.0 & 5.1 $\pm$ 1.8 & 1.0 $\pm$ 0.5 & 1.0 $\pm$ 0.6 \\
        3+4 & 7.2 $\pm$ 2.0 & 6.3 $\pm$ 2.0 & 1.1 $\pm$ 0.5 & 1.2 $\pm$ 0.6 \\
        5+6 & 8.5 $\pm$ 2.4 & 7.4 $\pm$ 2.2 & 1.2 $\pm$ 0.5 & 1.3 $\pm$ 0.5 \\
        \hline
        \end{tabular}
    \label{tab:en}
\end{table}

\section{Discussion and conclusions}\label{sec:disc}

We have observed the presence of a redshifted \lya\ emission component that, when observed through a narrow band, presents a bipolar morphology. This suggests that the nature of this component is a galactic-scale outflow of gas, likely a result of nuclear activity. The possibility of it being originated by cold inflowing gas, radiating in \lya\ through fluorescence, does not match the observed "hot" kinematic signatures (i.e., the width of $\sim 1,600$ \kms). The most realistic alternative hypothesis would be for \TXS\ to be undergoing a merger event with a close companion galaxy. And, although the velocity shift between the line centers is compatible with such a scenario, this would require an extremely hot galaxy (in terms of kinematics) with no detectable continuum emission. Additionally, the morphology extracted from the red tail of the profile yields a clearly extended profile with two bright spots close to the galactic center, much more compatible with outflowing gas stemming from the active nucleus of the host and voiding the central regions of gas. Moreover, the continuum emission is completely co-spatial with the main emission. All of these factors conflict with the companion and/or merger hypotheses.

The fact that the observed wide component is on the red side of the host emission, and not on the blue side as it is for outflows traced by nonresonant ionized lines (see \citealt{Concas_2019}), also falls within expectations. Outflow signatures in \lya\ have been an important topic of discussion in studies of radiation transfer: in particular, \cite{Verhamme_2006} predict the outflow signatures to appear mostly at positive relative velocities due to back-scattering and strong absorption at rest-frame wavelengths. In addition, the photons capable of traversing the intergalactic and intracluster medium without being absorbed are those that leave the galaxy with frequencies either close to the rest frame wavelength or redshifted. Conversely, "blue" \lya\ photons will be redshifted back to the central \lya\ wavelength as they travel; if they encounter neutral gas at that point, they are highly likely to be absorbed and consequently scattered. This phenomenon greatly hinders our ability to study the \lya\ emission at high velocities blueward of the systemic host emission, as any intervening cloud of neutral HI gas will scatter and further reduce the escape fraction of \lya\ radiation.

With all of these considerations in mind, we conclude that the most likely explanation for the observed emission profile is a galactic-scale ionized gas outflow. To quantitatively explore its characteristics, we estimated the mass of the outflow using the relation derived by \citet{osterbrock89}:

\begin{equation}
    M_{of}=7.5\,\times 10^{-3} \left(\frac{10^{4}}{n_e}\frac{L_{H\beta}}{L_\odot} \right) M_\odot
\end{equation}

assuming the ratios Ly$\alpha$/H$\alpha$=8.7 \citep{Sobral_2019} and H$\alpha$/H$\beta$=2.86, and a gas density of n$_e$ = 200 cm$^{-3}$. This last value was adopted to enable a comparison between our results and those obtained by \citet{Speranza_2021}. We derive the following physical quantities (see, e.g., \citealt{Fiore_2017}) for the outflow:\\
(i) a mass outflow rate:  $\dot{M}=3v_{out}\frac{M_{H\beta}}{R_{out}}= 9.3 M_\odot$ yr$^{-1}$, (ii) a kinetic energy: E$_{kin}=\frac{1}{2}M_{H\beta}v_{out}^{2}=10^{56.5}$ erg and (iii) a kinetic power  $\dot{E}_{kin}=\frac{1}{2}\dot{M}v_{out}^{2}= 10^{42.1}$ erg s$^{-1}$.

The jet power $P_{\rm jet} \sim 10^{47}$ erg s$^{-1}$, is a factor of $\sim10^5$ higher thnt the outflow kinetic power. In RGs at low redshifts, $z\lesssim 0.3$, \citet{Speranza_2021} measured outflow properties in the range $\dot{E}_{kin}=10^{42}-10^{45}$ erg s$^{-1}$ and estimated $P_{\rm jet}$ in the range $\sim 10^{44} - 3 \times 10^{45}$ erg s$^{-1}$. The typical ratio between the jet and outflow power is $\sim 0.1$. In \TXS\ this ratio is significantly lower, though similar to that observed in TN~J1049-1258, another HzRG from the same sample \citep{puga2023extended}, for which the observed ratio was $\dot{E}_{kin}/P_{jet}$=10$^{-4}$.

The radio source associated with TXS~0952-217 is unresolved, with a limit on its deconvolved size of $\lesssim 10$ kpc; with this and its steep radio spectrum, it fulfills the criteria for compact steep spectrum (CSS) sources (see, e.g., \citealt{odea21}). CSS sources, along with gigahertz peaked spectrum radio sources, have been associated with the earlier evolutionary phases of RGs (see \citealt{Fanti_95}, \citealt{Collier_2018}). The MUSE observations indicate that the outflow of ionized gas extends far beyond the radio structure. This is unusual, given that the expansion speed of the radio structure is expected to be higher than that of the outflow, in particular in the early phases of expansion where velocities as high as $\sim 0.1 \times c$ have been observed in CSS (e.g., \citealt{polatidis03}), while the projected outflow speed is just $\sim$ 500 \kms. Assuming a constant gas expansion speed, the lower bound for the age of the observed outflow is $\sim2\times10^{8}$ yr, much larger than the upper limit of the duration of radio activity, $3\times10^{5} v_{0.1}$ yr (where $v_{0.1}$ is the  expansion speed in $0.1\times c$ units). A possible way to reconcile these results is to envisage that the radio emission undergoes transitions between phases of high and low activity and that the current radio structure represents only the most recent period of a high activity state. 

Both the small size of the radio emission (when compared to the outflow size) and the lower ratio of outflow to jet kinetic power (compared to lower-redshift RGs) suggest that the relativistic jets and outflow are decoupled, and produced by two independent acceleration mechanisms: the jets might be driven by the magnetic field in the innermost regions of the accretion disk as expected if the jet launching process is due to the extraction of the rotational energy of the supermassive black hole \citep{blandford77} while the outflow is powered by radiation pressure from the outer regions of the accretion disk. A similar conclusion was proposed for low-z RGss, based on different grounds, by \citet{capetti23}.

The extension of the \lya\ nebula in \TXS\ is $\sim$24 kpc, smaller than other \lya\ nebulae around HzRGs at similar redshifts. \citet{Wang_2023} measured consistently larger nebular extensions for the other HzRGs observed by MUSE, ranging from $\sim 70$ to more than $\sim 300$ kpc. This difference can be explained by two factors. First, the observations used in that study are, on average, five to six times longer than our own. Consequently, they can detect zones of fainter nebular emission. Additionally, the authors acknowledge the possibility of a bias in their sample, which mostly includes well-known bright and extended sources. We will investigate this issue in more detail in a forthcoming paper in which we discuss the properties of all 16 targets of our MUSE observing project of HzRGs.

The assumption of co-spatiality between the AGN and the peak of \lya\ emission is not always correct; \citealt{Wang_2023} find a correlation between the eccentricity of \lya\ nebulae in HzRGs and the distance between the flux peak and the AGN position, obtained from radio or X-ray observations, where the distance increases as the nebulae become more asymmetric. In our case, the eccentricity measured via Gaussian fitting of the narrowband \lya\ image yields e=0.67$\pm$0.06, which would correspond to a very short distance ($\sim$1 kpc) between the \lya\ peak and the AGN position, assuming the relation holds true for the smaller, fainter HzRGs.

The observed \lya\ kinematics are within expectations with those observed in sources from the same sample of HzRGs. \citet{Wang_2023} observed W$_{80}$ values ranging from 10$^{3}$ to 3$\times$10$^{3}$\kms, which is entirely compatible with our measured value of W$_{80}$=1320$\pm$60\kms. Their study also observed a radial gradient for the line width, with the regions closest to the AGN position presenting the most disturbed kinematic signatures. This could not be quantifiably measured in our case, given that a radial width profile could not be performed with the number of available bins, but the most disturbed kinematics are coincident with the brightest spots of emission of the red component (see Figures \ref{fig:velmaps} and \ref{fig:lyatail}). This would, in principle, agree with the AGN being located in between these bright spots.

The presence of HI absorbers in the spectrum of \lya\ is well documented; recent studies have found very extended \lya\ nebulae with absorbing troughs up to hundreds of kiloparsecs away from the nucleus (\citealt{Silva_2017,Wang_2023}). The main absorber we observed, located very close to the systemic redshift, is detected in some of these studies and is predicted by radiative transfer simulations \citep{Verhamme_2006}. The presence of multiple absorbing sheets blueward of the central wavelength, originated from either cold ionized gas or expanding gas shells in the inter-stellar medium, is expected. Redshifted absorbers, however, are a rarer occurrence and could be the result of cold, infalling gas being accreted onto the galaxy. Additionally, given the faintness of CIV$\lambda$1550, we could not use its emission profile, which is also subject to scattering resonance, to test the possible enrichment of the observed HI clouds. A detailed analysis of the properties of these absorbers, which have shown a considerable degree of degeneracy in terms of their defining parameters (see \citealt{Silva_2017}), is beyond the scope of this work.

The observed line ratios are consistent with an AGN as the main photoionization mechanism. The value of \lya/HeII is well within the typical values measured in HzRGs \citep{Villar_Mart_n_2007_a} and other large \lya\ emitters, for example the Slug Nebula \citep{cantalupo19}. Additionally, the similarity between the values of \lya/HeII and \lya/CIV is also consistent with photoionization- as opposed to high-speed shocks, which could produce higher HeII fluxes \citep{allen08,arrigoni15,Travascio_2020}. The fact that both \lya/HeII and \lya/CIV are higher beyond 1$\sigma$ in the northwestern area, corresponding to bins 5 and 6, compared to the nuclear region (bin 1), could imply a slight change in the photoionization equilibrium, possibly due to increased star formation \citep{Villar_Mart_n_2007_a}. However, the values observed are still too low to sustain this claim.

For the two HzRGs discussed, \TN\ in \citet{puga2023extended} and \TXS\ throughout this study, the results of the MUSE observations point to the presence of an outflow. However, though they are apparently similar, they are in fact different in mant regards: \\
(i) their radio morphology (the former is an extended radio source, while the latter is compact), (ii) the geometry of their \lya\ emission (the former is one sided, while the latter is bipolar), (iii) their line emission (the latter presents high-ionization lines, while the former is only detected in \lya), and (iv) their gas kinematics (the former has a large velocity offset with small velocity dispersion while the latter has lower speed and a very large internal velocity spread).

It is worth noting that the marked difference in methodology utilized in these two studies might be the source of some of these differences.

Other studies of AGN-induced outflows in high-redshift galaxies differ from our analysis in both the observed outflow signatures and the interpretation of the physical processes behind them. For instance, a blue tail was found on the otherwise Gaussian profile of \lya\ on the radio-quiet quasar J1538+08 at z=3.6 \citep{Travascio_2020}. Some observations of this line are, apparently, wholly unaffected by absorption or scattering resonance effects, meaning there is no silver bullet and a case-by-case approach is still needed. Future studies will no doubt shed light on the differences behind the unabsorbed and the highly affected profiles observed in \lya.

The immediate surroundings of the \lya\-emitting gas and the ionization mechanism responsible for it are extremely important in determining the spectral shape and distribution of the emission, which means every object studied through this line needs specific considerations beyond what is required for nonresonant lines. Nonetheless, the amount of information it contains makes it one of the most interesting fields of research on ionized gas in distant objects.

The advent of high-resolution deep infrared spectroscopy, kick-started by the recent launch of JWST, will enable a number of these objects to be studied in their rest-frame optical window, where reliable tracers of ionized gas will allow for confirmation of the observed kinematics.

\vspace{5mm}

\begin{acknowledgements}
B.B. acknowledges financial support for the project “The MURALES project: exploring  AGN feedback in the most powerful radio-loud active galactic  nuclei” funded by PRIN INAF 2023 and based on observations collected at the European Organisation for Astronomical Research in the Southern Hemisphere under ESO programme 108.22FU.
CRA acknowledges the projects "Feeding and feedback in active galaxies”, with reference PID2019-106027GB-C42, funded by MICINN-AEI/10.13039/501100011033, and "Quantifying the impact of quasar feedback on galaxy evolution", with reference EUR2020-112266, funded by MICINN-AEI/10.13039/501100011033 and the European Union NextGenerationEU/PRTR.
G.V. acknowledges support from European Union’s HE ERC Starting Grant No. 101040227 - WINGS.
\end{acknowledgements}


\bibliography{aanda}{}

\begin{thebibliography}{76}
\expandafter\ifx\csname natexlab\endcsname\relax\def\natexlab#1{#1}\fi

\bibitem[{{Allen} {et~al.}(2008){Allen}, {Groves}, {Dopita}, {Sutherland}, \& {Kewley}}]{allen08}
{Allen}, M.~G., {Groves}, B.~A., {Dopita}, M.~A., {Sutherland}, R.~S., \& {Kewley}, L.~J. 2008, \apjs, 178, 20

\bibitem[{Andrae {et~al.}(2010)Andrae, Schulze-Hartung, \& Melchior}]{andrae2010dos}
Andrae, R., Schulze-Hartung, T., \& Melchior, P. 2010, Dos and don'ts of reduced chi-squared

\bibitem[{{Arrigoni Battaia} {et~al.}(2015){Arrigoni Battaia}, {Hennawi}, {Prochaska}, \& {Cantalupo}}]{arrigoni15}
{Arrigoni Battaia}, F., {Hennawi}, J.~F., {Prochaska}, J.~X., \& {Cantalupo}, S. 2015, \apj, 809, 163

\bibitem[{Arrigoni-Battaia {et~al.}(2018)Arrigoni-Battaia, Hennawi, Prochaska, Oñorbe, Farina, Cantalupo, \& Lusso}]{Arrigoni_Battaia_2018}
Arrigoni-Battaia, F., Hennawi, J.~F., Prochaska, J.~X., {et~al.} 2018, Monthly Notices of the Royal Astronomical Society, 482, 3162–3205

\bibitem[{Arrigoni~Battaia {et~al.}(2019)Arrigoni~Battaia, Obreja, Prochaska, Hennawi, Rahmani, Bañados, Farina, Cai, \& Man}]{Arrigoni_Battaia_2019}
Arrigoni~Battaia, F., Obreja, A., Prochaska, J.~X., {et~al.} 2019, Astronomy \& Astrophysics, 631, A18

\bibitem[{Balmaverde {et~al.}(2018)Balmaverde, Capetti, Marconi, \& Venturi}]{Balmaverde_2018}
Balmaverde, B., Capetti, A., Marconi, A., \& Venturi, G. 2018, Astronomy \& Astrophysics, 612, A19

\bibitem[{Behrens {et~al.}(2014)Behrens, Dijkstra, \& Niemeyer}]{Behrens_2014}
Behrens, C., Dijkstra, M., \& Niemeyer, J.~C. 2014, Astronomy \& Astrophysics, 563, A77

\bibitem[{Best {et~al.}(2006)Best, Kaiser, Heckman, \& Kauffmann}]{Best_2006}
Best, P.~N., Kaiser, C.~R., Heckman, T.~M., \& Kauffmann, G. 2006, Monthly Notices of the Royal Astronomical Society: Letters, 368, L67

\bibitem[{{Blandford} \& {Znajek}(1977)}]{blandford77}
{Blandford}, R.~D. \& {Znajek}, R.~L. 1977, \mnras, 179, 433

\bibitem[{{Borisova} {et~al.}(2016){Borisova}, {Cantalupo}, {Lilly}, {Marino}, {Gallego}, {Bacon}, {Blaizot}, {Bouch{\'e}}, {Brinchmann}, {Carollo}, {Caruana}, {Finley}, {Herenz}, {Richard}, {Schaye}, {Straka}, {Turner}, {Urrutia}, {Verhamme}, \& {Wisotzki}}]{borisova16}
{Borisova}, E., {Cantalupo}, S., {Lilly}, S.~J., {et~al.} 2016, \apj, 831, 39

\bibitem[{Brookes {et~al.}(2008)Brookes, Best, Peacock, Röttgering, \& Dunlop}]{Brookes_2008}
Brookes, M.~H., Best, P.~N., Peacock, J.~A., Röttgering, H. J.~A., \& Dunlop, J.~S. 2008, Monthly Notices of the Royal Astronomical Society, 385, 1297

\bibitem[{{Brookes} {et~al.}(2006){Brookes}, {Best}, {Rengelink}, \& {R{\"o}ttgering}}]{brookes06}
{Brookes}, M.~H., {Best}, P.~N., {Rengelink}, R., \& {R{\"o}ttgering}, H.~J.~A. 2006, \mnras, 366, 1265

\bibitem[{Cai {et~al.}(2019)Cai, Cantalupo, Prochaska, Arrigoni~Battaia, Burchett, Li, Chisholm, Bundy, \& Hennawi}]{Cai_2019}
Cai, Z., Cantalupo, S., Prochaska, J.~X., {et~al.} 2019, The Astrophysical Journal Supplement Series, 245, 23

\bibitem[{Cantalupo(2017)}]{Cantalupo_2017}
Cantalupo, S. 2017, in Gas Accretion onto Galaxies (Springer International Publishing), 195--220

\bibitem[{Cantalupo {et~al.}(2014)Cantalupo, Arrigoni-Battaia, Prochaska, Hennawi, \& Madau}]{Cantalupo_2014}
Cantalupo, S., Arrigoni-Battaia, F., Prochaska, J.~X., Hennawi, J.~F., \& Madau, P. 2014, Nature, 506, 63–66

\bibitem[{{Cantalupo} {et~al.}(2019){Cantalupo}, {Pezzulli}, {Lilly}, {Marino}, {Gallego}, {Schaye}, {Bacon}, {Feltre}, {Kollatschny}, {Nanayakkara}, {Richard}, {Wendt}, {Wisotzki}, \& {Prochaska}}]{cantalupo19}
{Cantalupo}, S., {Pezzulli}, G., {Lilly}, S.~J., {et~al.} 2019, \mnras, 483, 5188

\bibitem[{{Capetti} {et~al.}(2023){Capetti}, {Balmaverde}, {Baldi}, {Baum}, {Chiaberge}, {Grandi}, {Marconi}, {O'Dea}, \& {Venturi}}]{capetti23}
{Capetti}, A., {Balmaverde}, B., {Baldi}, R.~D., {et~al.} 2023, \aap, 671, A32

\bibitem[{{Cappellari} \& {Copin}(2003)}]{Cappellari2003}
{Cappellari}, M. \& {Copin}, Y. 2003, MNRAS, 342, 345

\bibitem[{Carniani {et~al.}(2016)Carniani, Marconi, Maiolino, Balmaverde, Brusa, Cano-Díaz, Cicone, Comastri, Cresci, Fiore, Feruglio, La~Franca, Mainieri, Mannucci, Nagao, Netzer, Piconcelli, Risaliti, Schneider, \& Shemmer}]{Carniani_2016}
Carniani, S., Marconi, A., Maiolino, R., {et~al.} 2016, Astronomy \& Astrophysics, 591, A28

\bibitem[{{Cavagnolo} {et~al.}(2010){Cavagnolo}, {McNamara}, {Nulsen}, {Carilli}, {Jones}, \& {B{\^\i}rzan}}]{Cavagnolo10}
{Cavagnolo}, K.~W., {McNamara}, B.~R., {Nulsen}, P.~E.~J., {et~al.} 2010, \apj, 720, 1066

\bibitem[{Collier {et~al.}(2018)Collier, Tingay, Callingham, Norris, Filipovi{\'{c} }, Galvin, Huynh, Intema, Marvil, O'Brien, Roper, Sirothia, Tothill, Bell, For, Gaensler, Hancock, Hindson, Hurley-Walker, Johnston-Hollitt, Kapi{\'{n}}ska, Lenc, Morgan, Procopio, Staveley-Smith, Wayth, Wu, Zheng, Heywood, \& Popping}]{Collier_2018}
Collier, J.~D., Tingay, S.~J., Callingham, J.~R., {et~al.} 2018, Monthly Notices of the Royal Astronomical Society

\bibitem[{{Coloma Puga} {et~al.}(2023){Coloma Puga}, Balmaverde, Capetti, Massaro, Almeida, Miley, Gilli, \& Marconi}]{puga2023extended}
{Coloma Puga}, M., Balmaverde, B., Capetti, A., {et~al.} 2023, An extended Lyman $\alpha$ outflow from a radio galaxy at z=3.7?

\bibitem[{Concas {et~al.}(2019)Concas, Popesso, Brusa, Mainieri, \& Thomas}]{Concas_2019}
Concas, A., Popesso, P., Brusa, M., Mainieri, V., \& Thomas, D. 2019, Astronomy \& Astrophysics, 622, A188

\bibitem[{Cresci {et~al.}(2015)Cresci, Mainieri, Brusa, Marconi, Perna, Mannucci, Piconcelli, Maiolino, Feruglio, Fiore, Bongiorno, Lanzuisi, Merloni, Schramm, Silverman, \& Civano}]{Cresci_2015}
Cresci, G., Mainieri, V., Brusa, M., {et~al.} 2015, The Astrophysical Journal, 799, 82

\bibitem[{De~Breuck {et~al.}(2010)De~Breuck, Seymour, Stern, Willner, Eisenhardt, Fazio, Galametz, Lacy, Rettura, Rocca-Volmerange, \& Vernet}]{De_Breuck_2010}
De~Breuck, C., Seymour, N., Stern, D., {et~al.} 2010, The Astrophysical Journal, 725, 36–62

\bibitem[{Donoso {et~al.}(2010)Donoso, Li, Kauffmann, Best, \& Heckman}]{Donoso_2010}
Donoso, E., Li, C., Kauffmann, G., Best, P.~N., \& Heckman, T.~M. 2010, Monthly Notices of the Royal Astronomical Society, 407, 1078–1089

\bibitem[{Dubois {et~al.}(2016)Dubois, Peirani, Pichon, {et~al.}}]{Dubois_2016}
Dubois, Y., Peirani, S., Pichon, C., {et~al.} 2016, Monthly Notices of the Royal Astronomical Society, 463, 3948

\bibitem[{Falder {et~al.}(2010)Falder, Stevens, Jarvis, Hardcastle, Lacy, McLure, Hatziminaoglou, Page, \& Richards}]{Falder_2010}
Falder, J.~T., Stevens, J.~A., Jarvis, M.~J., {et~al.} 2010, Monthly Notices of the Royal Astronomical Society

\bibitem[{{Fanti} {et~al.}(1995){Fanti}, {Fanti}, {Dallacasa}, {Schilizzi}, {Spencer}, \& {Stanghellini}}]{Fanti_95}
{Fanti}, C., {Fanti}, R., {Dallacasa}, D., {et~al.} 1995, Astronomy \& Astrophysics, 302, 317

\bibitem[{Finley {et~al.}(2014)Finley, Petitjean, Noterdaeme, \& Pâris}]{Finley_2014}
Finley, H., Petitjean, P., Noterdaeme, P., \& Pâris, I. 2014, A\&A, 572, A31

\bibitem[{Fiore {et~al.}(2017)Fiore, Feruglio, Shankar, Bischetti, Bongiorno, Brusa, Carniani, Cicone, Duras, Lamastra, Mainieri, Marconi, Menci, Maiolino, Piconcelli, Vietri, \& Zappacosta}]{Fiore_2017}
Fiore, F., Feruglio, C., Shankar, F., {et~al.} 2017, Astronomy \& Astrophysics, 601, A143

\bibitem[{Gronke {et~al.}(2015)Gronke, Bull, \& Dijkstra}]{Gronke_2015}
Gronke, M., Bull, P., \& Dijkstra, M. 2015, The Astrophysical Journal, 812, 123

\bibitem[{Gronke {et~al.}(2017)Gronke, Dijkstra, McCourt, \& Oh}]{Gronke_2017}
Gronke, M., Dijkstra, M., McCourt, M., \& Oh, S.~P. 2017, Astronomy \& Astrophysics, 607, A71

\bibitem[{Hayes(2015)}]{Hayes_2015}
Hayes, M. 2015, Publications of the Astronomical Society of Australia, 32

\bibitem[{Hogan {et~al.}(2015)Hogan, Edge, Hlavacek-Larrondo, Grainge, Hamer, Mahony, Russell, Fabian, McNamara, \& Wilman}]{Hogan_2015}
Hogan, M.~T., Edge, A.~C., Hlavacek-Larrondo, J., {et~al.} 2015, Monthly Notices of the Royal Astronomical Society, 453, 1201–1222

\bibitem[{{Hurley-Walker} {et~al.}(2017){Hurley-Walker}, {Callingham}, {Hancock}, {Franzen}, {Hindson}, {Kapi{\'n}ska}, {Morgan}, {Offringa}, {Wayth}, {Wu}, {Zheng}, {Murphy}, {Bell}, {Dwarakanath}, {For}, {Gaensler}, {Johnston-Hollitt}, {Lenc}, {Procopio}, {Staveley-Smith}, {Ekers}, {Bowman}, {Briggs}, {Cappallo}, {Deshpande}, {Greenhill}, {Hazelton}, {Kaplan}, {Lonsdale}, {McWhirter}, {Mitchell}, {Morales}, {Morgan}, {Oberoi}, {Ord}, {Prabu}, {Shankar}, {Srivani}, {Subrahmanyan}, {Tingay}, {Webster}, {Williams}, \& {Williams}}]{hurley17}
{Hurley-Walker}, N., {Callingham}, J.~R., {Hancock}, P.~J., {et~al.} 2017, \mnras, 464, 1146

\bibitem[{Jarvis {et~al.}(2003)Jarvis, Wilman, Rottgering, \& Binette}]{Jarvis_2003}
Jarvis, M.~J., Wilman, R.~J., Rottgering, H. J.~A., \& Binette, L. 2003, Monthly Notices of the Royal Astronomical Society, 338, 263–272

\bibitem[{Kale {et~al.}(2015)Kale, Venturi, Cassano, Giacintucci, Bardelli, Dallacasa, \& Zucca}]{Kale_2015}
Kale, R., Venturi, T., Cassano, R., {et~al.} 2015, Astronomy \& Astrophysics, 581, A23

\bibitem[{Kormendy \& Ho(2013)}]{Kormendy_2013}
Kormendy, J. \& Ho, L.~C. 2013, Annual Review of Astronomy and Astrophysics, 51, 511

\bibitem[{Lacy {et~al.}(2020)Lacy, Baum, Chandler, Chatterjee, Clarke, Deustua, English, Farnes, Gaensler, Gugliucci, Hallinan, Kent, Kimball, Law, Lazio, Marvil, Mao, Medlin, Mooley, Murphy, Myers, Osten, Richards, Rosolowsky, Rudnick, Schinzel, Sivakoff, Sjouwerman, Taylor, White, Wrobel, Andernach, Beasley, Berger, Bhatnager, Birkinshaw, Bower, Brandt, Brown, Burke-Spolaor, Butler, Comerford, Demorest, Fu, Giacintucci, Golap, Güth, Hales, Hiriart, Hodge, Horesh, Ivezi{\'{c}}, Jarvis, Kamble, Kassim, Liu, Loinard, Lyons, Masters, Mezcua, Moellenbrock, Mroczkowski, Nyland, O'Dea, O'Sullivan, Peters, Radford, Rao, Robnett, Salcido, Shen, Sobotka, Witz, Vaccari, van Weeren, Vargas, Williams, \& Yoon}]{Lacy_2020}
Lacy, M., Baum, S.~A., Chandler, C.~J., {et~al.} 2020, Publications of the Astronomical Society of the Pacific, 132, 035001

\bibitem[{Lau {et~al.}(2022)Lau, Hamann, Gillette, Perrotta, Rupke, Wylezalek, \& Zakamska}]{Lau_2022}
Lau, M.~W., Hamann, F., Gillette, J., {et~al.} 2022, Monthly Notices of the Royal Astronomical Society, 515, 1624–1643

\bibitem[{{Madau} \& {Dickinson}(2014)}]{Madau2014}
{Madau}, P. \& {Dickinson}, M. 2014, \araa, 52, 415

\bibitem[{{Miley} \& {De Breuck}(2008)}]{Miley_2008}
{Miley}, G. \& {De Breuck}, C. 2008, Astronomy and Astrophysics Review, 15, 67

\bibitem[{{O'Dea} \& {Saikia}(2021)}]{odea21}
{O'Dea}, C.~P. \& {Saikia}, D.~J. 2021, \aapr, 29, 3

\bibitem[{{Osterbrock}(1989)}]{osterbrock89}
{Osterbrock}, D.~E. 1989, {Astrophysics of gaseous nebulae and active galactic nuclei} (University Science Books)

\bibitem[{{Polatidis} \& {Conway}(2003)}]{polatidis03}
{Polatidis}, A.~G. \& {Conway}, J.~E. 2003, \pasa, 20, 69

\bibitem[{Price-Whelan {et~al.}(2018)Price-Whelan, Sipőcz, Günther, {et~al.}}]{astropy:2018}
Price-Whelan, A.~M., Sipőcz, B.~M., Günther, H.~M., {et~al.} 2018, Astronomical Journal, 156, 123

\bibitem[{Prochaska {et~al.}(2013)Prochaska, Hennawi, Lee, Cantalupo, Bovy, Djorgovski, Ellison, Lau, Martin, Myers, Rubin, \& Simcoe}]{Prochaska_2013}
Prochaska, J.~X., Hennawi, J.~F., Lee, K.-G., {et~al.} 2013, The Astrophysical Journal, 776, 136

\bibitem[{{Ramos Almeida} \& Ricci(2017)}]{almeida2017nuclear}
{Ramos Almeida}, C. \& Ricci, C. 2017, Nuclear obscuration in active galactic nuclei

\bibitem[{{Reuland} {et~al.}(2003){Reuland}, {van Breugel}, {R{\"o}ttgering}, {de Vries}, {Stanford}, {Dey}, {Lacy}, {Bland-Hawthorn}, {Dopita}, \& {Miley}}]{Reuland_2003}
{Reuland}, M., {van Breugel}, W., {R{\"o}ttgering}, H., {et~al.} 2003, \apj, 592, 755

\bibitem[{Rigby {et~al.}(2013)Rigby, Hatch, Röttgering, Sibthorpe, Chiang, Overzier, Herbonnet, Borgani, Clements, Dannerbauer, De~Breuck, De~Lucia, Kurk, Maschietto, Miley, Saro, Seymour, \& Venemans}]{Rigby_2013}
Rigby, E.~E., Hatch, N.~A., Röttgering, H. J.~A., {et~al.} 2013, Monthly Notices of the Royal Astronomical Society, 437, 1882

\bibitem[{Robitaille {et~al.}(2013)Robitaille, Tollerud, Greenfield, {et~al.}}]{astropy:2013}
Robitaille, T.~P., Tollerud, E.~J., Greenfield, P., {et~al.} 2013, Astronomy \& Astrophysics, 558, A33

\bibitem[{Rocca-Volmerange {et~al.}(2004)Rocca-Volmerange, Le~Borgne, De~Breuck, Fioc, \& Moy}]{Rocca_Volmerange_2004}
Rocca-Volmerange, B., Le~Borgne, D., De~Breuck, C., Fioc, M., \& Moy, E. 2004, Astronomy \& Astrophysics, 415, 931–940

\bibitem[{Schawinski {et~al.}(2015)Schawinski, Koss, Berney, \& Sartori}]{Schawinski_2015}
Schawinski, K., Koss, M., Berney, S., \& Sartori, L.~F. 2015, Monthly Notices of the Royal Astronomical Society, 451, 2517

\bibitem[{Scholtz {et~al.}(2020)Scholtz, Harrison, Rosario, Alexander, Chen, Kakkad, Mainieri, Tiley, Turner, Cirasuolo, Sharples, \& Stach}]{Scholtz_2020}
Scholtz, J., Harrison, C.~M., Rosario, D.~J., {et~al.} 2020, Monthly Notices of the Royal Astronomical Society, 492, 3194–3216

\bibitem[{Scholtz {et~al.}(2021)Scholtz, Harrison, Rosario, Alexander, Knudsen, Stanley, Chen, Kakkad, Mainieri, \& Mullaney}]{Scholtz_2021}
Scholtz, J., Harrison, C.~M., Rosario, D.~J., {et~al.} 2021, Monthly Notices of the Royal Astronomical Society, 505, 5469–5487

\bibitem[{Seymour {et~al.}(2007)Seymour, Stern, De~Breuck, Vernet, Rettura, Dickinson, Dey, Eisenhardt, Fosbury, \& Lacy}]{Seymour_2007}
Seymour, N., Stern, D., De~Breuck, C., {et~al.} 2007, The Astrophysical Journal Supplement Series, 171

\bibitem[{Silva {et~al.}(2017)Silva, Humphrey, Lagos, Villar-Mart{\'{\i} }n, Morais, di~Serego~Alighieri, Cimatti, Fosbury, Overzier, Vernet, \& Binette}]{Silva_2017}
Silva, M., Humphrey, A., Lagos, P., {et~al.} 2017, Monthly Notices of the Royal Astronomical Society, 474, 3649

\bibitem[{Sobral \& Matthee(2019)}]{Sobral_2019}
Sobral, D. \& Matthee, J. 2019, Astronomy \& Astrophysics, 623, A157

\bibitem[{Speranza {et~al.}(2021)Speranza, Balmaverde, Capetti, {et~al.}}]{Speranza_2021}
Speranza, G., Balmaverde, B., Capetti, A., {et~al.} 2021, Astronomy \& Astrophysics, 653, A150

\bibitem[{{Steidel} {et~al.}(2010){Steidel}, {Erb}, {Shapley}, {Pettini}, {Reddy}, {Bogosavljevi{\'c}}, {Rudie}, \& {Rakic}}]{steidel2010}
{Steidel}, C.~C., {Erb}, D.~K., {Shapley}, A.~E., {et~al.} 2010, \apj, 717, 289

\bibitem[{Swinbank {et~al.}(2015)Swinbank, Vernet, Smail, De~Breuck, Bacon, Contini, Richard, Röttgering, Urrutia, \& Venemans}]{Swinbank_2015}
Swinbank, A.~M., Vernet, J. D.~R., Smail, I., {et~al.} 2015, Monthly Notices of the Royal Astronomical Society, 449, 1298–1308

\bibitem[{Travascio {et~al.}(2020)Travascio, Zappacosta, Cantalupo, Piconcelli, Battaia, Ginolfi, Bischetti, Vietri, Bongiorno, D'Odorico, Duras, Feruglio, Vignali, \& Fiore}]{Travascio_2020}
Travascio, A., Zappacosta, L., Cantalupo, S., {et~al.} 2020, Astronomy \& Astrophysics, 635, A157

\bibitem[{Uchiyama {et~al.}(2022)Uchiyama, Yamashita, Toshikawa, Kashikawa, Ichikawa, Kubo, Ito, Kawakatu, Nagao, Toba, Ono, Harikane, Imanishi, Kajisawa, Lee, \& Liang}]{Uchiyama_2022}
Uchiyama, H., Yamashita, T., Toshikawa, J., {et~al.} 2022, The Astrophysical Journal, 926, 76

\bibitem[{van Ojik {et~al.}(1996{\natexlab{a}})van Ojik, Röttgering, Carilli, Miley, Bremer, \& Macchetto}]{vanojik1996radio}
van Ojik, R., Röttgering, H. J.~A., Carilli, C.~L., {et~al.} 1996{\natexlab{a}}, A radio galaxy at z=3.6 in a giant rotating Lyman $\alpha$ halo

\bibitem[{van Ojik {et~al.}(1996{\natexlab{b}})van Ojik, Röttgering, Miley, \& Hunstead}]{vanojik1996gaseous}
van Ojik, R., Röttgering, H. J.~A., Miley, G.~K., \& Hunstead, R.~W. 1996{\natexlab{b}}, The Gaseous Environments of Radio Galaxies in the Early Universe: Kinematics of the Lyman $\alpha$ Emission

\bibitem[{Venemans {et~al.}(2006)Venemans, Röttgering, Miley, van Breugel, De~Breuck, Kurk, Pentericci, Stanford, Overzier, Croft, \& Ford}]{Venemans_2006}
Venemans, B.~P., Röttgering, H. J.~A., Miley, G.~K., {et~al.} 2006, Astronomy \& Astrophysics, 461, 823–845

\bibitem[{Verhamme {et~al.}(2006)Verhamme, Schaerer, \& Maselli}]{Verhamme_2006}
Verhamme, A., Schaerer, D., \& Maselli, A. 2006, Astronomy \& Astrophysics, 460, 397

\bibitem[{Villar-Martin {et~al.}(2007b)Villar-Martin, Sanchez, Humphrey, Dijkstra, Di~Serego~Alighieri, De~Breuck, \& Gonzalez~Delgado}]{Villar_Martin_2007_b}
Villar-Martin, M., Sanchez, S.~F., Humphrey, A., {et~al.} 2007b, Monthly Notices of the Royal Astronomical Society, 378, 416–428

\bibitem[{Villar-Martín {et~al.}(2007a)Villar-Martín, Humphrey, De~Breuck, Fosbury, Binette, \& Vernet}]{Villar_Mart_n_2007_a}
Villar-Martín, M., Humphrey, A., De~Breuck, C., {et~al.} 2007a, Monthly Notices of the Royal Astronomical Society, 375, 1299–1310

\bibitem[{Wang {et~al.}(2024)Wang, Wylezalek, Breuck, Vernet, Rupke, Zakamska, Vayner, Lehnert, Nesvadba, \& Stern}]{wang2024jwst}
Wang, W., Wylezalek, D., Breuck, C.~D., {et~al.} 2024, JWST discovers an AGN ionization cone but only weak radiative-driven feedback in a powerful $z$$\approx$3.5 radio-loud AGN

\bibitem[{Wang {et~al.}(2021)Wang, Wylezalek, De~Breuck, Vernet, Humphrey, Villar~Martín, Lehnert, \& Kolwa}]{Wang_2021}
Wang, W., Wylezalek, D., De~Breuck, C., {et~al.} 2021, Astronomy \& Astrophysics, 654, A88

\bibitem[{Wang {et~al.}(2023)Wang, Wylezalek, Vernet, De~Breuck, Gullberg, Swinbank, Villar~Martín, Lehnert, Drouart, Arrigoni~Battaia, Humphrey, Noirot, Kolwa, Seymour, \& Lagos}]{Wang_2023}
Wang, W., Wylezalek, D., Vernet, J., {et~al.} 2023, Astronomy \& Astrophysics, 680, A70

\bibitem[{{Weilbacher} {et~al.}(2020){Weilbacher}, {Palsa}, {Streicher}, {Bacon}, {Urrutia}, {Wisotzki}, {Conseil}, {Husemann}, {Jarno}, {Kelz}, {P{\'e}contal-Rousset}, {Richard}, {Roth}, {Selman}, \& {Vernet}}]{Weilbacher20}
{Weilbacher}, P.~M., {Palsa}, R., {Streicher}, O., {et~al.} 2020, \aap, 641, A28

\bibitem[{{Willott} {et~al.}(1999){Willott}, {Rawlings}, {Blundell}, \& {Lacy}}]{willott99}
{Willott}, C.~J., {Rawlings}, S., {Blundell}, K.~M., \& {Lacy}, M. 1999, \mnras, 309, 1017

\bibitem[{Wylezalek {et~al.}(2013)Wylezalek, Galametz, Stern, Vernet, Breuck, Seymour, Brodwin, Eisenhardt, Gonzalez, Hatch, Jarvis, Rettura, Stanford, \& Stevens}]{Wylezalek_2013}
Wylezalek, D., Galametz, A., Stern, D., {et~al.} 2013, The Astrophysical Journal, 769, 79

\end{thebibliography}
\bibliographystyle{aa}

\end{document}